\documentstyle[twocolumn,aps]{revtex}
\begin{document}
\draft
\title{Non-Relativistic Fermions Coupled to Transverse Gauge-Fields:
The Single-Particle Green's Function in Arbitrary Dimension}
\author{Peter Kopietz}
\address{
Institut f\H{u}r Theoretische Physik der Universit\H{a}t G\H{o}ttingen,\\
Bunsenstr.9, D-37073 G\H{o}ttingen, Germany}
\date{\today}
\maketitle
\begin{abstract}
We use a bosonization approach
to calculate the single-particle  Green's function $G ( {\bf{r}} , \tau )$ of
non-relativistic fermions
coupled to transverse gauge-fields in arbitrary
dimension $d$. We find that
in $d>3$ transverse gauge-fields do not destroy the Fermi liquid, although
for $d < 6$ the quasi-particle damping is anomalously large.
For $d \rightarrow 3$
the quasi-particle residue vanishes as
$Z \propto \exp [ - \frac{1}{2 \pi ( d-3)} ( \frac{ \kappa}{mc } )^2 ]$, where
$\kappa$ is the Thomas-Fermi wave-vector, $m$ is the mass of the electrons, and
$c$ is the velocity of the gauge-particle.
In $d=3$ the system is a Luttinger liquid,
with anomalous dimension $\gamma_{\bot} = \frac{1}{6 \pi} ( \frac{ \kappa}{mc}
)^2$.
For $d < 3$ we find that $G ( {\bf{r}} , 0 )$ decays exponentially
at large distances.

\end{abstract}
\narrowtext

More than 20 years ago
Holstein, Norton and Pincus\cite{Holstein73}
realized that the self-energy of non-relativistic fermions that are coupled to
the transverse
radiation field is plagued by logarithmic singularities.
They concluded that
in $d=3$ dimensions such a system cannot be a Fermi liquid.
However, due to the smallness of the fine-structure  constant
$ \frac{e^2}{c } \approx \frac{1}{137}$ the deviations from
conventional Fermi-liquid behavior are of little practical consequences.
The discovery of unusual normal-state properties in the
high-temperature superconductors\cite{Ioffe89} and recent
investigations of  half-filled quantum
Hall systems\cite{Halperin93} have revived the interest in this problem.
In these systems
the relevant dimensionless
coupling constant may not be small,
and the effective dimensionality is not necessarily $d=3$.
Thus we are lead to consider the problem of studying
electrons that are minimally coupled
to a gauge-field $A_{\mu} = [ \phi , {\bf{A}} ]$.
In an Euclidean functional integral approach, the model is defined
in terms of an action
 $S = S_{1} \left\{ \psi \right\} + S_{2} \left\{ \psi , A_{\mu} \right\}
 + S_{3} \left\{ A_{\mu} \right\}$,
with
 \begin{eqnarray}
 S_{1} \left\{ \psi \right\}
 & = & \beta \sum_{ k } \psi^{\dagger}_{ k } \left[ -i \tilde{\omega}_{n} +
 \xi_{\bf{k}}  \right] \psi_{ k }
 \; \; \; ,
 \label{eq:S1Fourier}
 \\
 S_{2} \left\{ \psi , A_{\mu} \right\}
 & = & g
 \sum_{ q } \left[
 i \rho_{q}  \phi_{ -q}
 - {\bf{j}}_{ q}  \cdot {\bf{A}}_{ -q}
 \right]
 \; \; \; ,
 \label{eq:S2Fourier}
 \\
 S_{3} \left\{ {A}_{\mu}  \right\} & = &
 \frac{1}{2} \sum_{q} \left[ \tilde{f}_{ {\bf{q}} }^{-1} \phi_{-q} \phi_{q}
 + \tilde{h}_{q}^{-1} {\bf{A}}_{-q} \cdot {\bf{A}}_{q} \right]
 \label{eq:S3Fourier}
 \; \; \; .
 \end{eqnarray}
Here $S_{1}  \{ \psi \}$ is the Euclidean action
of non-relativistic spinless electrons, which are represented by a
Grassmann-field with Fourier components $\psi_{k}$.
The excitation energy relative to the chemical potential $\mu$ is
$\xi_{\bf{k}} = \frac{ {\bf{k}}^2}{2m} - \mu $,
where $m$ is the mass of the electrons.
The coupling between matter and gauge-field is described by
$S_{2} \{ \psi , A_{\mu} \}$,
where $g$ is the relevant coupling constant.
The composite Grassmann-fields
 $\rho_{q} = \sum_{k} \psi^{\dagger}_{k} \psi_{k+q}$
 and
${\bf{j}}_{ q }$
represent density and gauge-invariant current density.
Finally, $S_{3} \{ A_{\mu } \}$ is the action for the
gauge-field {\it{in Coulomb gauge}}, ${\bf{q}} \cdot {\bf{A}}_{q} = 0$.
For the Maxwell-field the functions $\tilde{f}_{\bf{q}}$ and $\tilde{h}_{q}$
are
 \begin{equation}
 \tilde{f}_{\bf{q}} = \frac{ \beta}{{V}} \frac{ 4 \pi }{ {\bf{q}}^2 }
 \; \; \; , \; \; \;
 \tilde{h}_{{q}} = \frac{ \beta}{{V}} \frac{ 4 \pi }{
 {\bf{q}}^2  + ( \frac{ \omega_{m} }{  c } )^2 }
 \; \; \; ,
 \label{eq:tildefhdef}
 \end{equation}
where $V$ is the volume of the system, $\beta$ is the inverse temperature,
and $c$ is the velocity of the
gauge-particle.
Throughout this work we shall
use the convention that $k = [ {\bf{k}}, i \tilde{\omega}_{n} ]$
and $q = [ {\bf{q}} , i \omega_{m} ]$,
where
the fermionic frequencies are $\tilde{\omega}_{n} = 2 \pi ( n + \frac{1}{2} ) /
\beta$,
and the bosonic ones are $\omega_{m} = 2 \pi m / \beta $.

The integration over the transverse gauge-field  gives rise to a retarded
current-current interaction
between the matter degrees of freedom\cite{Holstein73,Pethick89}.
In $d \leq 3$ this interaction cannot be treated within the usual perturbative
many-body theory,
so that it is necessary to attack the problem by means of
a non-perturbative approach,
such as renormalization group and scaling
methods\cite{Gan93}-\cite{Chakravarty95},
generalized $\frac{1}{N}$ expansions\cite{Ioffe94},
higher dimensional bosonization methods\cite{Kwon94}, or other
infinite resummation schemes\cite{Khveshchenko93,Castellani94}.

The bosonization technique in $d> 1$
is based on ideas due to Luther\cite{Luther79}.
It was rediscovered by Haldane\cite{Haldane92},
and has been further developed in Refs.\cite{Houghton93,Castro94}.
More recently,
Sch\H{o}nhammer and the present author\cite{Kopietz94}
have developed a powerful formulation of  higher-dimensional bosonization
which is based on functional integration.
In the literature on one-dimensional bosonization these functional
techniques have been known for some time\cite{Fogedby76},
but have not received much attention.
In Ref.\cite{Kopietz95} we used
our approach to
calculate corrections
to the non-interacting boson approximation, which exist even in
$d=1$ if the energy dispersion is not linearized.
Up until now these corrections could not be calculated by means of the
conventional operator approach\cite{Houghton93,Castro94}.
In the present work we shall show that
the problem of electrons coupled to transverse gauge-fields
can be treated within our method in a very simple
and elegant way.

We start by partitioning
the Fermi surface into patches of area $\Lambda^{d-1}$
and subdividing the degrees of freedom close to the
Fermi surface into squat boxes $K^{\alpha}$ of hight $\lambda$ associated
with the patches\cite{Haldane92,Houghton93,Kopietz94,Kopietz95}.
The boxes and patches are
labelled by an index $\alpha$ in some convenient ordering.
In Ref.\cite{Kopietz94}  we have decoupled the
two-body density-density interaction
via a generalized Hubbard-Stratonowich
field $\phi^{\alpha}_{q}$ which couples to the
local density-field $\rho^{\alpha}_{q} =
\sum_{k} \Theta^{\alpha} ( {\bf{k}} ) \psi^{\dagger}_{k} \psi_{k+q}$,
where $\Theta^{\alpha} ( {\bf{k}} )$ is unity if the wave-vector ${\bf{k}}$
lies inside the box $K^{\alpha}$, and vanishes otherwise.
Note that by construction
$\rho_{q} = \sum_{\alpha} \rho^{\alpha}_{q}$.
Imagine now that in our gauge action defined in
Eqs.\ref{eq:S1Fourier}-\ref{eq:S3Fourier}
we integrate first over the $\phi$-field, decompose then the total density
into its contributions from the various patches and finally  decouple again the
matter action by means of the auxiliary field $\phi^{\alpha}_{q}$
introduced in Refs.\cite{Kopietz94,Kopietz95}.
Clearly the longitudinal part of the
new decoupled action can be obtained from
Eqs.\ref{eq:S2Fourier} and \ref{eq:S3Fourier}
by replacing
$\sum_{q}  \rho_{q} \phi_{-q} \rightarrow
\sum_{q} \sum_{\alpha}   \rho_{q}^{\alpha} \phi_{-q}^{\alpha}$ and
$\sum_{q} \tilde{f}_{q}^{-1} \phi_{-q} \phi_{q} \rightarrow
\sum_{q}  \sum_{\alpha \alpha^{\prime}}
 [ \underline{\tilde{f}}_{q}^{-1}]^{\alpha \alpha^{\prime} }
 \phi^{\alpha}_{-q} \phi^{\alpha^{\prime}}_{q} $,
where $\underline{ \tilde{f}}_{q} $ is a matrix in the patch labels,
with matrix elements
$[ \underline{ \tilde{f}}_{q} ]^{\alpha \alpha^{\prime}} =
\tilde{f}_{\bf{q}}$\cite{footnote1}.
Hence, the Hubbard-Stratonowich field $\phi^{\alpha}_{q}$ that was formally
introduced in
Ref.\cite{Kopietz94} to decouple the two-body density interaction
can be identified physically with
the patch-version of the longitudinal gauge-field.
Evidently the transverse part of the gauge-field can be treated in exactly the
same way,
so that the correlation functions of the $\psi$-field
can be obtained by functional integration over the
patch gauge-field action
 $\tilde{S} = {S} \left\{ \psi \right\} + \tilde{S}_{2} \left\{ \psi ,
A^{\alpha}_{\mu} \right\}
 + \tilde{S}_{3} \left\{ A^{\alpha}_{\mu} \right\}$.
The form of $\tilde{S}_{2}$ and $\tilde{S}_{3}$ is similar to
Eqs.\ref{eq:S2Fourier} and
\ref{eq:S3Fourier}, the only difference being that the gauge-fields and the
densities carry an extra patch-label $\alpha$.
Thus, the second terms in Eqs.\ref{eq:S2Fourier} and \ref{eq:S3Fourier} are
replaced by
 $\sum_{ q } \sum_{\alpha}
  {\bf{j}}^{\alpha}_{ q}  \cdot {\bf{A}}^{\alpha}_{ -q}$ and
 $
 \sum_{q}
 \sum_{\alpha \alpha^{\prime}}
 [ \underline{\tilde{h}}_{q}^{-1}]^{\alpha \alpha^{\prime}}
 {\bf{A}}_{-q}^{\alpha} \cdot {\bf{A}}_{q}^{\alpha^{\prime}}$, where
$\underline{ \tilde{h}}_{q} $ is a matrix in the patch labels,
with matrix elements
$[ \underline{ \tilde{h}}_{q} ]^{\alpha \alpha^{\prime}} =
\tilde{h}_{{q}}$\cite{footnote1}.
The patch current density is
${\bf{j}}_{ q }^{\alpha} = {\bf{j}}^{para, \alpha}_{q} + {\bf{j}}^{dia,
\alpha}_{q}$
 with ${\bf{j}}^{para, \alpha}_{q}  =
 \sum_{k} \Theta^{\alpha} ( {\bf{k}} ) \frac{  ( {\bf{k}} + {\bf{q}}/ 2 ) }{m c
}
 \psi^{\dagger}_{k} \psi_{k+q}$ and
 ${\bf{j}}^{dia , \alpha}_{ q}  =
 - \frac{ g}{ 2 m c^2 \beta} \sum_{ q^{\prime}}
 {\bf{A}}^{\alpha}_{ q-q^{\prime} } \rho^{\alpha}_{q^{\prime}} $.

To calculate the
fermionic Green's function $G ( k )$,
we start from the
representation of $G ( k )$
as a functional integral over the fields $\psi$ and
$A^{\alpha}_{\mu}$, and integrate first over the $\psi$-field.
In this way we obtain
the Green's function ${\cal{G}} ( k , k^{\prime} ; \left\{ A^{\alpha}_{\mu }
\right\} )$ of the
matter field for a frozen configuration of the gauge-field. Note that
this function is not diagonal in momentum-energy space, because
the gauge fields act like a dynamical random-potential,
which breaks the translational invariance in space and time.
The full fermionic Green's function can then be obtained by averaging
${\cal{G}} ( k , k ; \left\{ A^{\alpha}_{\mu } \right\} )$
over all configurations of the gauge fields,
 \begin{equation}
 G ( k ) = \left<
 {\cal{G}} ( k , k ; \left\{ A^{\alpha}_{\mu } \right\}  ) \right>_{A}
 \label{eq:avgauge}
 \; \; \; ,
 \end{equation}
where $< \ldots >_{A}$ denotes functional average with
the effective gauge-field action $\tilde{S}_{gf} \left\{ A^{\alpha}_{\mu}
\right\}$ that
contains in addition to $\tilde{S}_{3} \{ A^{\alpha}_{\mu } \}$ a contribution
$\tilde{S}_{kin} \{ A^{\alpha}_{\mu } \}$ describing the influence of the
matter
degrees of freedom on the dynamics of the gauge-field.
Note that ${\cal{G}} ( k , k ; \{ A^{\alpha}_{\mu } \} ) = [ \hat{G}]_{kk}$,
where
the infinite matrix $\hat{G}$ is defined by writing
$S_{1} \{ \psi \} + \tilde{S}_{2} \{ \psi , A^{\alpha}_{\mu} \}
= - \beta \sum_{k k^{\prime} } \psi^{\dagger}_{k} [  \hat{G}^{-1} ]_{k
k^{\prime}}
\psi_{k^{\prime}}$.
Of course, in general we cannot invert the matrix $\hat{G}^{-1}$.
However, if the Thomas-Fermi wave-vector
$\kappa = ( 4 \pi g^2 \nu )^{\frac{1}{2}}$
is small compared with $k_{F}$, then
the matrix $\hat{G}^{-1}$ is dominated by
a narrow band of width $\kappa \ll k_{F}$ in the wave-vector label.
Here $\nu$ is the density of states
at the Fermi energy, and $k_{F} $ is the Fermi wave-vector.
If we then partition $\hat{G}^{-1}$ into block-matrices
corresponding to the patches and choose
the cutoffs $\Lambda$ and $\lambda$ large compared with $\kappa$,
it is obvious that $\hat{G}^{-1}$
is approximately block-diagonal,
$[ \hat{G}^{-1} ]_{k k^{\prime}} = \sum_{\alpha}
\Theta^{\alpha} ( {\bf{k}} )
\Theta^{\alpha} ( {\bf{k}}^{\prime} )
[ ( \hat{G}^{\alpha})^{-1} ]_{k k^{\prime}}$,
with diagonal blocks $ ( \hat{G}^{\alpha} )^{-1}$
associated with the boxes $K^{\alpha}$.
Note that scattering processes which transfer momentum
between different $K^{\alpha}$ are neglected in this approximation,
so that possible non-trivial effects due to the {\it{global}} topology of the
Fermi surface are not taken into account\cite{Khlebnikov94}.

The crucial point is now that {\it{within a given
block}} we may linearize the energy dispersion,
$\xi_{ {\bf{k}}^{\alpha} + {\bf{q}} } \approx {\bf{v}}^{\alpha} \cdot
{\bf{q}}$,
where ${\bf{k}}^{\alpha}$ is a vector on the Fermi surface that points
to the suitably defined center of $K^{\alpha}$.
Although each diagonal block $( \hat{G}^{\alpha} )^{-1}$ is still an infinite
matrix in frequency space,
it can then be inverted exactly by means of a method originally due to
Schwinger\cite{Schwinger62}.
Thus,
the matrix elements $[ \hat{G}^{\alpha}] _{kk}$
can be {\it{explicitly calculated
as functionals of the $A^{\alpha}_{\mu}$}}.
Finally we average this expression over all configurations of the
gauge-field, using the Gaussian approximation for the effective gauge-field
action.
As shown in Ref.\cite{Kopietz95}, in a model with linearized energy dispersion
the Gaussian approximation becomes very accurate at high densities due to a
large-scale cancellation between self-energy and vertex corrections
(generalized closed loop
theorem).
The Coulomb gauge condition ${\bf{q}} \cdot {\bf{A}}^{\alpha}_{q}$ is easily
imposed in our functional integral approach via the
Fadeev-Popov method\cite{Sterman93}.
It should be kept in mind, however, that the single-particle Green's function
is gauge-dependent,
although physical observables derived from it are not.
The final result for the interacting Green's function is
 \begin{eqnarray}
 G (k)  & = &  \sum_{\alpha} \Theta^{\alpha} ( {\bf{k}} )
 \int d {\bf{r}} \int_{0}^{\beta} d \tau
 e^{ - i [ ( {\bf{k}} - {\bf{k}}^{\alpha} ) \cdot  {\bf{r}}
 - \tilde{\omega}_{n}  \tau  ] }
 \nonumber
 \\
 & \times &
 {{G}}^{\alpha}_{0} ( {\bf{r}}  , \tau  )
 e^{ Q^{\alpha}_{\|} ( {\bf{r}} , \tau )
 + Q^{\alpha}_{\bot} ( {\bf{r}} , \tau ) }
 \; \; \; ,
 \label{eq:Gkres2}
 \end{eqnarray}
 \begin{eqnarray}
 Q^{\alpha}_{\|}
 ( {\bf{r}} , \tau ) & = &
 R^{\alpha}_{\|}
 - S^{\alpha}_{\|} ( {\bf{r}} , \tau )
  \; \;   , \;  \;
 R^{\alpha}_{\|} =  \lim_{ {\bf{r}} , \tau \rightarrow 0} S^{\alpha}_{\|}
({\bf{r}} , \tau )
 \;  ,
 \label{eq:Qlondef}
 \\
 {Q}^{\alpha}_{\bot}
 ( {\bf{r}} , \tau )  & =  &
 {R}^{\alpha}_{\bot}
 - {S}^{\alpha}_{\bot} ( {\bf{r}} , \tau )
  \; \;   , \;  \;
 {R}^{\alpha}_{\bot} =  \lim_{ {\bf{r}} , \tau \rightarrow 0}
{S}^{\alpha}_{\bot} ({\bf{r}} , \tau )
 \;  \; ,
 \label{eq:Qraddef}
 \end{eqnarray}
 \begin{eqnarray}
 S^{\alpha}_{\|}
 ( {\bf{r}}  , \tau   )
  & = &
 \frac{1}{\beta {{V}}} \sum_{ q }
 \frac{   f^{RPA, \alpha}_{q}
  \cos ( {\bf{q}} \cdot  {\bf{r}}
  - {\omega}_{m}  \tau  )
 }
 {
 ( i \omega_{m} - {\bf{v}}^{\alpha} \cdot {\bf{q}} )^{2 }}
 \; \; \; ,
  \label{eq:Slondef}
  \\
 {S}^{\alpha}_{\bot}
 ( {\bf{r}}  , \tau   )
  & = &
 \frac{1}{\beta {{V}}} \sum_{ q }
 \frac{   h_{q}^{RPA, \alpha}
  \cos ( {\bf{q}} \cdot  {\bf{r}}
  - {\omega}_{m}  \tau  )
 }
 {
 ( i \omega_{m} - {\bf{v}}^{\alpha} \cdot {\bf{q}} )^{2 }}
 \label{eq:Sraddef}
 \; \; \; ,
 \end{eqnarray}
 \begin{equation}
 G^{\alpha}_{0} ( {\bf{r}}  , \tau  )
 =
 \delta^{(d-1)}_{\Lambda} ( {\bf{r}}_{\bot}  )
 \left( \frac{ - i}{2 \pi} \right)
 \frac{1}
 {
  r_{\|}
 + i | {\bf{v}}^{\alpha} |  \tau }
 \; \; \; ,
 \label{eq:Gpatchreal1}
 \end{equation}
 where $r_{\|} = {\bf{r}} \cdot \hat{\bf{v}}^{\alpha}$,
 with
 $\hat{\bf{v}}^{\alpha} = {\bf{v}}^{\alpha} / | {\bf{v}}^{\alpha} |$.
 For length scales
 $| {\bf{r}}_{\bot} | \gg \Lambda^{-1}$ the
 function $\delta^{(d-1)}_{\Lambda} ( {\bf{r}}_{\bot} )$
 can be treated as a $d-1$-dimensional Dirac-$\delta$ function of the
components
 of $ {\bf{r}} $
 orthogonal to $\hat{\bf{v}}^{\alpha}$.
Here $f^{RPA , \alpha}_{q}$ is the effective screened
interaction due to the
longitudinal component of the gauge-field,
as defined in Refs.\cite{Kopietz94,Kopietz95}. The term $Q^{\alpha}_{\|} (
{\bf{r}} , \tau )$
has been discussed extensively in our previous work\cite{Kopietz94}. Here we
focus on the
contribution $Q^{\alpha}_{\bot} ( {\bf{r}} , \tau )$ due to the transverse
part of the gauge field.
In Ref.\cite{Kopietz95b} we shall give an expression for
$h^{RPA, \alpha}_{q}$
for an arbitrary shape of the Fermi surface
and general patch-dependent bare interactions $[
\underline{\tilde{h}}_{q}]^{\alpha \alpha^{\prime}}$.
For simplicity, in this discussion
we shall restrict ourselves to a spherical
%$d$-dimensional
Fermi surface
(so that  $| {\bf{v}}^{\alpha} |
= v_{F} $ for all $\alpha$) and
constant interaction matrices
$ [ \underline{\tilde{h}}_{q} ]^{\alpha \alpha^{\prime}} = \tilde{h}_{q}$,
where $\tilde{h}_{q}$ is
given in Eq.\ref{eq:tildefhdef}. In this case
 \begin{equation}
  h^{RPA,\alpha}_{q} = -   \frac{v_{F}^2}{c^2 \nu  }
 \frac{ 1 - ({\hat{\bf{v}}}^{\alpha} \cdot {\hat{\bf{q}}} )^2}
 { \left( \frac{{\bf{q}} }{ \kappa } \right)^2 +
 \left( \frac{v_{F}}{c} \right)^2 \left[
 ( \frac{\omega_{m} }{ v_{F} \kappa  } )^2 +
 \bar{g}_{d} \left( \frac{i \omega_{m} }{ v_{F} | {\bf{q}} | } \right)
 \right] }
 \label{eq:hrpasphere}
 \; \; \; ,
 \end{equation}
where $\hat{\bf{q}} = {\bf{q}} / | {\bf{q}} | $, and the function $\bar{g}_{d}
( z )$ is given by
 \begin{equation}
 \bar{g}_{d} ( z ) = -  {\gamma}_{d}  z  \int_{0}^{\pi}
 \frac{ d \theta}{\pi} \frac{ \sin^d \theta }{ \cos \theta - z }
 \; \; \; .
 \label{eq:gddef}
 \end{equation}
Here ${\gamma}_{d} = \frac{\pi}{d-1} \frac{ \Omega_{d-1}}{\Omega_{d}}$,
where $\Omega_{d} = { 2 \pi^{\frac{d}{2}}}/ { \Gamma ( \frac{d}{2} ) }$ is the
surface of the $d$-dimensional unit sphere.
Note that ${\gamma}_{d} \rightarrow \frac{\pi}{2}$ for $d \rightarrow 1$.
Because $G^{\alpha}_{0} ( {\bf{r}} , \tau )$
is proportional to $\delta_{\Lambda}^{(d-1)} ( {\bf{r}}_{\bot} )$, we may
set ${\bf{r}} = r_{\|} \hat{\bf{v}}^{\alpha}$ in the rest of the integrand
of Eq.\ref{eq:Gkres2}.
Furthermore,
close to and below three dimensions the value of the integral in
Eq.\ref{eq:Sraddef}
is dominated by the regime $ \frac{ | \omega_{m} | }{v_{F} | {\bf{q}} | } \ll
1$,
so that we may approximate in Eq.\ref{eq:gddef}
 \begin{equation}
 \bar{g}_{d} \left( \frac{ i  \omega_{m}}{v_{F} | {\bf{q}} | } \right) \approx
{\gamma}_{d}
 \frac{ | \omega_{m} |}{ v_{F} | {\bf{q}} | }
 \label{eq:denomleading}
 \; \; \; .
 \end{equation}
Note that to this order it is consistent to
neglect the term of order $\omega_{m}^2$ in the denominator
of Eq.\ref{eq:hrpasphere}.
After some trivial rescalings we obtain in the limit $V, \beta \rightarrow
\infty$
 \begin{eqnarray}
 Q_{\bot}^{\alpha} ( r_{\|} \hat{ {\bf{v}}}^{\alpha} ,  \tau ) &  =  &
  -  C_{d}
  \left( \frac{\kappa}{mc} \right)^{d-1}
 \int_{0}^{\pi} d \theta ( \sin \theta )^d | \cos \theta |^{\frac{d-5}{2}}
 \nonumber
 \\
 &  & \hspace{-24mm} \times
 \int_{0}^{\infty} d p {p}^{d-2}
 \int_{- \infty}^{\infty} { d u}
 \frac{ 1 - \cos \left[ |
 \cos \theta |^{\frac{3}{2}} p \kappa_{d} (  {r}_{\|} - u s_{\theta} v_{F}
{{\tau}} ) \right] }
 { \left[ p^2 + | u | \right]
 \left[ iu - s_{\theta}  \right]^2 }
 \nonumber
 \\
 & &
 \label{eq:Qbot5}
 \end{eqnarray}
where
$C_{d} = \frac{d-1}{2 \pi^2} \left( {\gamma}_{d} \right)^{\frac{d-1}{2}}$
is a numerical constant,
$\kappa_{d} =  {\gamma}_{d}^{\frac{1}{2}}   \frac{c  }{v_{F}} \kappa $
is the relevant wave-vector scale for the transverse gauge-field, and
$s_{\theta} = sign ( \cos \theta )$.

{}From now on we shall restrict ourselves to ${\tau} = 0$.
The $u$-integration in Eq.\ref{eq:Qbot5} is then easily carried out,
 \begin{eqnarray}
 Q_{\bot}^{\alpha} ( r_{\|} \hat{ {\bf{v}}}^{\alpha} ,  0 )   & =  &
  -  C_{d}
  \left( \frac{\kappa}{mc} \right)^{d-1}
  2 \int_{0}^{\infty} d p {p}^{d-2}
  \nonumber
  \\
  &  & \hspace{-15mm} \times
 \frac{ \pi p^2 - p^4 - 1 + ( p^4- 1 ) \ln p^2 }{[ 1 + p^4]^2}
 A_{d} \left(
  p \kappa_{d} | {r}_{\|} | \right)
 \label{eq:Qbotr2}
 \; \; \; ,
 \end{eqnarray}
where the dimensionless function $A_{d} ( x )$ is given by
 \begin{equation}
 A_{d} ( x )  =
 \int_{0}^{\pi} d \theta ( \sin \theta )^d | \cos \theta |^{\frac{d-5}{2}}
 \left[ 1 - \cos \left(
 | \cos \theta |^{\frac{3}{2}} x \right) \right]
 \label{eq:addef}
 \; \; .
 \end{equation}
Obviously the large-$r_{\|}$ asymptotics of
 $Q_{\bot}^{\alpha} ( r_{\|} \hat{ {\bf{v}}}^{\alpha} ,  0 )  $ is determined
by
the behavior of $A_{d} ( x )$ as $|x| \rightarrow \infty$.
In $d > 3 $
the limit
 $A_{d}^{\infty} \equiv \lim_{x \rightarrow \infty} A_{d} ( x )$
exists. For $d<6$ the leading correction is
 $A_{d} ( x ) \sim A_{d}^{\infty} - {A_{d}^{\prime}} |x|^{- \frac{d-3}{3} }$,
with
 \begin{equation}
 A_{d}^{\infty}   =
 \frac{ \Gamma ( \frac{ d+1}{2} ) \Gamma ( \frac{ d-3}{4} ) }{  \Gamma ( \frac{
3d -1 }{4} ) }
 \; \; , \; \;
 A_{d}^{\prime}
  =
 \frac{4}{3} \Gamma ( \frac{ d-3}{3} )
 \sin ( \frac{ \pi d }{6} )
 \; .
 \label{eq:cdres}
 \end{equation}
It follows that for
  $ \kappa_{d} | r_{\|} | \rightarrow \infty$
 \begin{equation}
 Q_{\bot}^{\alpha} ( r_{\|} \hat{ {\bf{v}}}^{\alpha} ,  0 )     \sim
  - a_{d}
  \left( \frac{\kappa}{mc } \right)^{d-1}
  = R_{\bot}^{\alpha}
  \; \; \; ,
 \label{eq:Rradres}
 \end{equation}
with $a_{d}
 =  A_{d}^{\infty} B_{d} C_{d}$, where
 \begin{equation}
 B_{d}  =
  2 \int_{0}^{\infty} d p {p}^{d-2}
 \frac{ \pi p^2 - p^4 - 1 + ( p^4- 1 ) \ln p^2 }{[ 1 + p^4]^2}
 \label{eq:bddef}
 \; \; \; .
 \end{equation}
The integral is convergent for $ 1 < d < 5$ and can be evaluated
analytically\cite{Kopietz95b}.
For $d=3$ we obtain $B_{3} = \frac{1}{2}$.

Since $R^{\alpha}_{\bot}$ is finite,
transverse gauge-fields do
not destroy the Fermi-liquid in $d > 3$: their contribution
to the quasi-particle residue
is $Z^{\alpha}_{\bot} = e^{R^{\alpha}_{\bot}}$.
However, in
$d < 6$ the damping is anomalously large, since the
leading correction
decays slower than $|x|^{-1}$\cite{Kopietz94}.
In the limit $d \rightarrow 3$,
$C_{3} = \frac{1}{4 \pi }$ and
$A_{d}^{\infty} \sim \frac{4}{d-3} $,
so that
 \begin{equation}
 Z_{\bot}^{\alpha} = e^{R^{\alpha}_{\bot}} \propto
 e^{ -  \frac{ 1 }{ 2 \pi (d-3)  }
 ( \frac{ \kappa}{mc} )^2 }
 \label{eq:Zradres1}
 \; \; \; , \; \; \; 0 < d-3 \ll 1
 \; \; \; .
 \end{equation}

In the {\it{marginal}} case $d=3$ it is easy to show that
 $A_{3} ( x ) \sim \frac{4}{3}  \ln ( | x |) $ for
$|x| \rightarrow \infty$,
so that for $   \kappa_{3} | r_{\|} |  \rightarrow \infty$
 \begin{equation}
 Q_{\bot}^{\alpha} ( r_{\|} \hat{ {\bf{v}}}^{\alpha} ,  0 ) \sim
  -  \frac{4}{3} B_{3} C_{3}
  \left( \frac{\kappa}{mc} \right)^{2}
  \ln \left( \kappa_{3} | {r}_{\|}| \right)
  \; \; \; .
  \label{eq:Qbotrd3}
  \end{equation}
Exponentiating this expression  yields
 \begin{equation}
 e^{Q^{\alpha}_{\bot} ( r_{\|} \hat{\bf{v}}^{\alpha}  , 0 ) }
 \propto   \left( \kappa_{3} {| r_{\|} | }  \right)^{ - \gamma_{\bot} }
 \; \; \; , \; \; \;
  \kappa_{3} | r_{\|}|  \rightarrow  \infty
 \label{eq:eQrad3}
 \; \; \; ,
 \end{equation}
where the {\it{anomalous dimension}} is
 $\gamma_{\bot}  =  \frac{1}{6 \pi }
  \left( \frac{\kappa}{mc } \right)^{2}$.
In the case of electromagnetism at realistic densities $\frac{ \kappa}{mc}
\approx 10^{-2}$, so that
$ \gamma_{\bot} \approx 5 \times 10^{-6}$.
Although this is extremely small,
at distances large compared with
$\kappa_{3}^{-1}$ the true asymptotic behavior
of the Green's function is characterized by a non-universal power law.
This is precisely the Luttinger liquid behavior familiar from
one-dimensional systems with regular interactions.

In $d < 3$ the function $A_{d} ( x )$ diverges
for $|x| \rightarrow \infty$
stronger than logarithmic,
$A_{d} ( x ) \sim {A}_{d}^{\prime \prime} | x |^{\frac{3-d}{3} }  $
with ${A}_{d}^{\prime \prime}  =    \frac{4}{3-d} \Gamma ( \frac{d}{3} )
 \sin ( \frac{ \pi d}{6} ) $.
It follows that for $ \kappa_{d} | {r}_{\|}  |  \rightarrow  \infty$
 \begin{equation}
 Q_{\bot}^{\alpha} ( r_{\|} \hat{ {\bf{v}}}^{\alpha} ,  0 ) \sim
  - \tilde{a}_{d}
  \left( \frac{\kappa}{mc} \right)^{d-1}
  \left( \kappa_{d} | r_{\|} |  \right)^{ \frac{3-d}{3} }
  \; \; \; ,
  \label{eq:Qbotrd}
  \end{equation}
with
 $\tilde{a}_{d} = A_{d}^{\prime \prime} B_{ d + \frac{3}{2} } C_{d}$.
Thus, for $d< 3$ the Green's function
decays at equal times and large distances faster than any power.
Obviously
the quasi-particle picture is not valid but the non-universal
power laws characteristic for conventional Luttinger liquids
do not appear.
Such a system can be called a {\it{generalized Luttinger liquid.}}

We would like to emphasize that our expression for
$Q_{\bot}^{\alpha} ( {\bf{r}} , \tau )$  in Eq.\ref{eq:Qbot5} and the
subsequent
equations derived from it are independent of the
box cutoffs $\lambda$ and $\Lambda$.
This is due to the fact that
$\kappa$ is the relevant ultraviolet cutoff for ${\bf{q}}$-integrations.
Because we have assumed
$\kappa \ll  \lambda , \Lambda  \ll k_{F}$,
the integrals do not depend on the precise choice of $\lambda$ and $\Lambda$.
In this regime the bosonization approach is most powerful
and leads to cutoff-independent results
that involve only physical quantities.
On the other hand, for
$ \lambda , \Lambda  \ll \kappa $
the ${\bf{q}}$-integrations extend over the entire volume of the
boxes $K^{\alpha}$,
so that the cutoff $\lambda$ appears
explicitly in the final expression for the Green's function\cite{Kwon94}.
The marginality of $d=3$ is also evident in
renormalization group calculations\cite{Gan93,Chakravarty95}, but
our bosonization approach goes beyond these works,
because the calculation of the full Green's function in arbitrary $d$
has been reduced to a well defined purely mathematical problem:
After substituting Eq.\ref{eq:Qbot5} into  Eq.\ref{eq:Gkres2},
the Fourier transformation can in principle be carried out numerically, or
perhaps even by some sophisticated analytical method.
Finally, it should be mentioned that in Refs.\cite{Ioffe94,Castellani94} the
validity of the bosonization approach
in the gauge-field problem has been questioned.
Note, however, that
the generalized closed loop theorem\cite{Kopietz95}
guarantees that the cancellations between self-energy and vertex corrections,
which are responsible for the accuracy of the Gaussian approximation at high
densities,
happen also in the case of gauge-fields\cite{Kopietz95b}.
Whether non-linearities in the energy dispersion
modify the asymptotics of the Green's function
is currently under investigation.
We have preliminary evidence that this is not the case\cite{Kopietz95b}.

In summary, we calculated the Green's function
of electrons coupled to the Maxwell-field at high densities.
In the absence of spontaneous symmetry breaking
the system is a Fermi liquid in $d > 3$, a Luttinger
liquid in $d=3$, and a generalized Luttinger liquid in $d < 3$.
Our non-perturbative method is very general and
can be used to study the coupling between electrons and any
other bosonic field, such as phonons or magnons\cite{Kopietz95b}.

I am grateful to K. Sch\H{o}nhammer for many discussions, fruitful
collaborations,
and comments on the manuscript.
I would also like to thank G. E. Castilla for detailed comments on the
manuscript and
W. Metzner for discussions.

%              R E F E R E N C E S
%


\begin{thebibliography}{99}
\bibitem{footnote1}
Of course, a matrix with constant elements is not invertible.
However, the appearance of $\underline{ \tilde{f}}_{q}^{-1}$
and $\underline{ \tilde{h}}_{q}^{-1}$
is only an  intermediate step.
Formally, we assume that we have regularized
$\underline{\tilde{f}}_{q}^{-1}$ and $\underline{ \tilde{h}}_{q}^{-1}$
in some convenient way and remove
the regularization at the end of the calculation.
\end{thebibliography}
\end{document}